\RequirePackage{fix-cm}
\documentclass[12pt]{article}
\usepackage[numbers]{natbib}
\usepackage[utf8]{inputenc}

\usepackage{times}%

\usepackage{natbib}
\usepackage{amsmath}
\usepackage{dcolumn}
\usepackage{endnotes}
\usepackage{graphics}
\usepackage{graphicx}
\usepackage{caption}
\usepackage{subcaption}
\usepackage{multirow}

\usepackage[table]{xcolor}
\definecolor{lightgray}{gray}{0.96}

\usepackage[affil-it]{authblk}
\usepackage{tikz}
\usetikzlibrary{shadows,calc}
\def\shadowshift{3pt,-3pt}
\def\shadowradius{6pt}

\colorlet{innercolor}{black!60}
\colorlet{outercolor}{gray!05}

\newcommand\drawshadow[1]{
    \begin{pgfonlayer}{shadow}
        \shade[outercolor,inner color=innercolor,outer color=outercolor] ($(#1.south west)+(\shadowshift)+(\shadowradius/2,\shadowradius/2)$) circle (\shadowradius);
        \shade[outercolor,inner color=innercolor,outer color=outercolor] ($(#1.north west)+(\shadowshift)+(\shadowradius/2,-\shadowradius/2)$) circle (\shadowradius);
        \shade[outercolor,inner color=innercolor,outer color=outercolor] ($(#1.south east)+(\shadowshift)+(-\shadowradius/2,\shadowradius/2)$) circle (\shadowradius);
        \shade[outercolor,inner color=innercolor,outer color=outercolor] ($(#1.north east)+(\shadowshift)+(-\shadowradius/2,-\shadowradius/2)$) circle (\shadowradius);
        \shade[top color=innercolor,bottom color=outercolor] ($(#1.south west)+(\shadowshift)+(\shadowradius/2,-\shadowradius/2)$) rectangle ($(#1.south east)+(\shadowshift)+(-\shadowradius/2,\shadowradius/2)$);
        \shade[left color=innercolor,right color=outercolor] ($(#1.south east)+(\shadowshift)+(-\shadowradius/2,\shadowradius/2)$) rectangle ($(#1.north east)+(\shadowshift)+(\shadowradius/2,-\shadowradius/2)$);
        \shade[bottom color=innercolor,top color=outercolor] ($(#1.north west)+(\shadowshift)+(\shadowradius/2,-\shadowradius/2)$) rectangle ($(#1.north east)+(\shadowshift)+(-\shadowradius/2,\shadowradius/2)$);
        \shade[outercolor,right color=innercolor,left color=outercolor] ($(#1.south west)+(\shadowshift)+(-\shadowradius/2,\shadowradius/2)$) rectangle ($(#1.north west)+(\shadowshift)+(\shadowradius/2,-\shadowradius/2)$);
        \filldraw ($(#1.south west)+(\shadowshift)+(\shadowradius/2,\shadowradius/2)$) rectangle ($(#1.north east)+(\shadowshift)-(\shadowradius/2,\shadowradius/2)$);
    \end{pgfonlayer}
}

\pgfdeclarelayer{shadow} 
\pgfsetlayers{shadow,main}

\newsavebox\mybox
\newlength\mylen

\newcommand\shadowimage[2][]{%
\setbox0=\hbox{\includegraphics[#1]{#2}}
\setlength\mylen{\wd0}
\ifnum\mylen<\ht0
\setlength\mylen{\ht0}
\fi
\divide \mylen by 120
\def\shadowshift{\mylen,-\mylen}
\def\shadowradius{\the\dimexpr\mylen+\mylen+\mylen\relax}
\begin{tikzpicture}
\node[anchor=south west,inner sep=0] (image) at (0,0) {\includegraphics[#1]{#2}};
\drawshadow{image}
\end{tikzpicture}}

\begin{document}

\title{An Open Solution to Provide Personalized Feedback for Building Energy Management}


\author{Andrea Monacchi, Fabio Versolatto, Manuel Herold, Dominik Egarter, Andrea M. Tonello, Wilfried Elmenreich%
\thanks{A. Monacchi, M. Herold, D. Egarter and W. Elmenreich are with the Institute of Networked and Embedded Systems, Alpen-Adria-Universität Klagenfurt,
F. Versolatto and A. M. Tonello are with WiTiKee Srl}}




\maketitle

\begin{abstract}
The integration of renewable energy sources increases the complexity in mantaining the power grid.
In particular, the highly dynamic nature of generation and consumption demands for a better utilization of energy resources,
which seen the cost of storage infrastructure, can only be achieved through demand-response.
Accordingly, the availability of energy and potential overload situations can be reflected using a price signal.
The effectiveness of this mechanism arises from the flexibility of device operation, which is nevertheless  
heavily reliant on the exchange of information between the grid and its consumers.
In this paper, we investigate the capability of an interactive energy management system to timely inform users on energy usage, in order to promote an optimal use of local resources.
In particular, we analyze data being collected in several households in Italy and Austria to gain insights into usage behavior and drive the design of more effective systems.
The outcome is the formulation of energy efficiency policies for residential buildings, as well as the design of an energy management system, consisting of hardware measurement units and a management software. 
The Mjölnir framework, which we release for open use, provides a platform where various feedback concepts can be implemented and assessed.
This includes widgets displaying disaggregated and aggregated consumption information, as well as daily production and tailored advices.
The formulated policies were implemented as an advisor widget able to autonomously analyze usage and provide tailored energy feedback.
%
\end{abstract}

\section{Introduction}\label{sec:introduction}
The progressive installation of renewable energy generators and the diffusion of electric vehicles contribute to destabilizing the offer and demand of energy in the grid.
Demand-side management can compensate this problem by exploiting a bidirectional information channel between utilities and customers to balance the demand to available supply.
This includes both promotion of efficiency and conservation~\cite{monacchi:2013Nov}, as well as direct scheduling of loads to off-peak periods~\cite{palensky}.
Thus, efficiency can generally be achieved by i) replacing devices with more efficient ones, ii) improving the efficiency of the building (e.g., using a better insulation), and iii) optimizing energy usage.
A first possibility is to analyze energy utilization to improve the overall efficiency.
This is normally implemented through energy audits, which can take place as surveys and interviews.
Nevertheless, the availability of high resolution smart meter data can allow companies for remote data analysis.
In \cite{beckel2013class}, traces from more than 3000 households are used to extract specific customers' properties.
Dynamic pricing schemes can offer an incentive to operate loads when the demand and cost of energy is lower.
To fully reflect energy availability, schemes should dynamically consider the offer of energy, 
so that users can allocate the energy necessary for their activities by bidding an amount in real-time.
The main problem with classic billing mechanisms is the delay between the feedback and the actual energy usage, which makes an understanding of energy use completely impossible.
A first possibility to increase the information resolution is to rely on smart metering.
Prepaid billing is another possibility, and was shown leading to average savings of 11\% in UK, regardless of disconnections from the grid~\cite{prepaid}.
Providing energy utilization feedback makes users aware of the energy necessary to operate devices. 
A possibility to provide unobtrusive feedback is the use of ambient interfaces, such as the power-aware cord \cite{Gustafsson:2005}.

\begin{figure}[h]\centering
	\includegraphics[trim=2.22cm 19.2cm 2.9cm 2cm,clip,width=0.9\columnwidth]{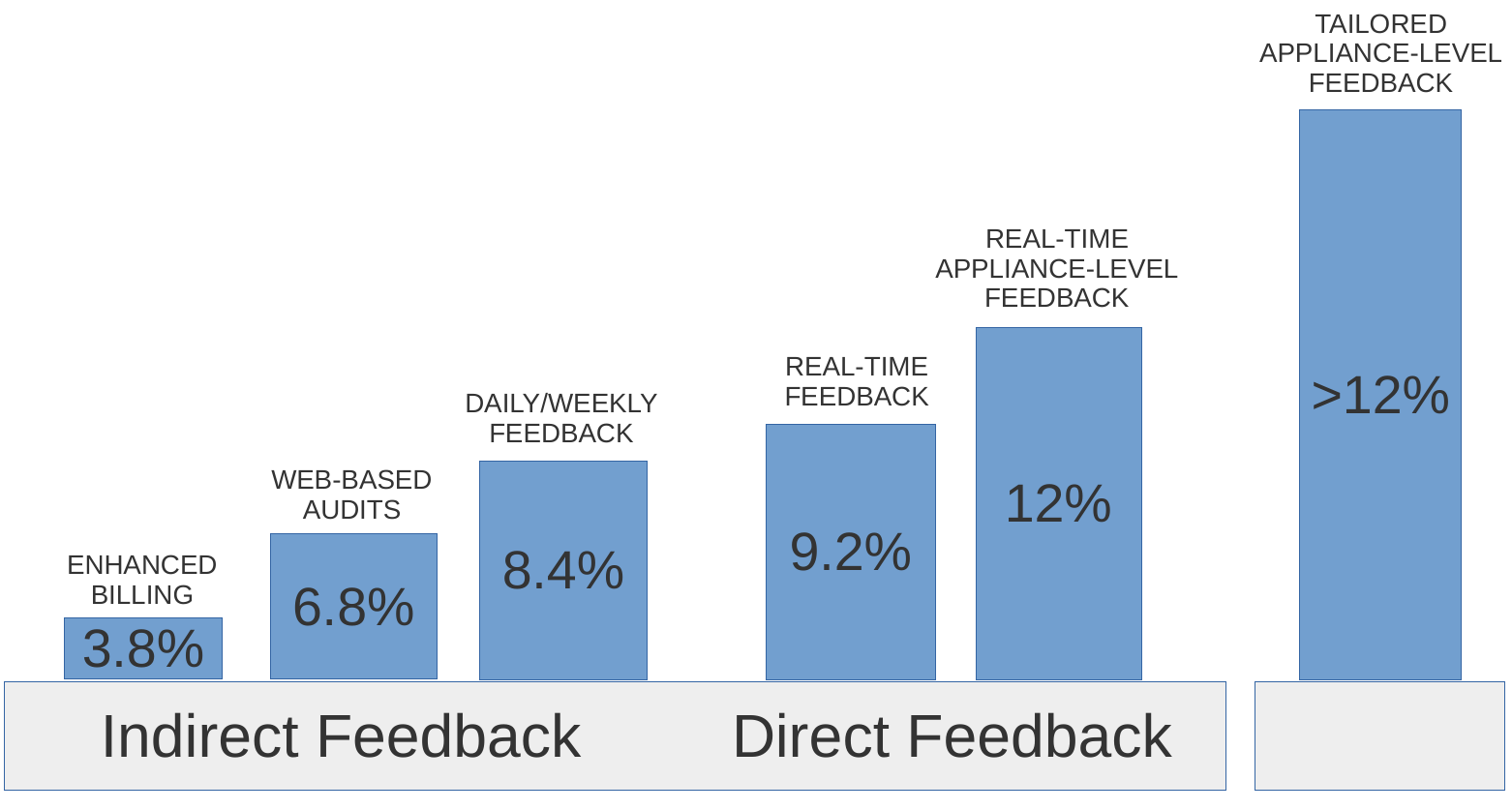}
\caption{Effectiveness of feedback \cite{aceee,holygrail}}\label{img:chart}\end{figure}

Darby~\cite{darby} classifies feedback in two categories:
\begin{itemize}
  \item \textit{Indirect}, when it provides consumption information after it occurred 
  \item \textit{Direct}, when the feedback concerns the amount of energy in use
\end{itemize}
Darby also shows that real-time consumption information can effectively raise user awareness, leading to a reduction in energy uses of up to 15\%.
On the other hand, indirect information is necessary to enable learning mechanisms, and consequently, long-term change.
%
%
Similarly, \cite{Bonino2012383} identifies \textit{antecedent} and \textit{consequent} strategies.
Antecedent strategies aim at preventing certain behaviors, for instance using goal-setting and advices, 
while consequent strategies concern direct and indirect feedback, which also includes monetary and social rewarding.
However, studies have also shown that in spite of awareness, the effectiveness of these systems in making people responsible depends on their sensitivity and motivation \cite{Strengers:2011}.
The analysis in \cite{aceee} relies on 36 studies carried out between 1995 and 2010 to show that consumption information at device level can lead the highest energy savings (see Fig. \ref{img:chart}).
Moreover, feedback mechanisms were shown more effective when exploiting a user model to offer personalized services, such as advices, with estimated savings of around 20\% \cite{holygrail}. 
Another possibility is to recommend use of particularly greedy devices in cheaper tariff periods, such as in the AgentSwitch~\cite{eps349815}. 
The evaluation carried out on 10 users for 3 months showed the system actually finding cheaper tariffs for most of users \cite{eps346991}.
In spite of the large amount of work already done for assessing different feedback mechanisms, an open solution where concepts can be implemented and assessed is still missing.
Therefore, such an energy management system should:
\begin{itemize}
  \item \textit{Use a non-technical measure to quantify energy utilization.}	
  Energy price offers a simple non-technical measure unit for energy, although expressing use in terms of costs might not be effective because of the very little amounts \cite{Bonino2012383}.
  \item \textit{Appliance-level information along with overall running costs.} 
  Fine-grained consumption information was shown leading to higher savings, and can be more effective to understand expenses and promote replacement with more efficient ones \cite{holygrail}.
  \item \textit{Direct feedback.} 
  In particular, real-time feedback was shown more effective than pure historical data \cite{aceee}.
  \item \textit{User-specific energy usage.} 
  The availability of a wallet for each resident would help tracking individual expenses and enable self-learning.
  For instance, this might be used by parents in limiting the time their kids spend on certain appliances (e.g., video game consoles). 
  \item \textit{Exploit user models.} 
  The availability of fine-grained consumption information allows for building usage models, which can be used to compare users and return tailored advices, such as device curtailment and replacement. 
\end{itemize}
This paper analyzes energy usage data collected in Austria and Italy to derive energy efficiency policies, with the ultimate goal of designing an open energy management system.
The remainder of this paper is organized as follows: 
Section~\ref{sec:monergy} overviews the results of MONERGY, a project that explores the use of Home Energy Management Systems (HEMS) to foster alteration of consumption patterns and achieve better utilization of local resources.
Previously, we identified common electrical devices and we have analyzed the penetration of renewable energy \cite{monacchi:2013Nov}, as well as users' attitude towards smart homes \cite{tamer:smarthome}.
A measurement campaign was then carried out and is described in Section~\ref{sec:monergy}.
Section~\ref{sec:analysis} reports an extensive analysis of consumption patterns for the monitored environments.
The main outcome is the formulation of energy efficiency policies in Sect.~\ref{sec:policies}.
Based on our findings, we propose in Section~\ref{sec:platform_solution} a system that monitors energy usage and provides users with tailored feedback, based on open hardware and source components.
In particular, section~\ref{sec:advisor} introduces Mjölnir, a modular energy dashboard providing widgets to display disaggregated and aggregated consumption information, as well as daily production and tailored advices.
We expect the tool to be beneficial for both individual users and researchers carrying out field tests on persuasive technologies, whose contribution is particularly encouraged.
The energy efficiency policies were implemented as a specific widget, which is described in Sect.~\ref{sec:advices}.
An estimation of their impact in terms of savings is provided in Sect.~\ref{sec:evaluation}.
Section~\ref{sec:conclusions} concludes the paper, summarizing the contribution and anticipating future developments.

\section{Residential Energy Management in Austria and Italy}\label{sec:monergy}
In the MONERGY\footnote{http://www.monergy-project.eu} project we have explored solutions to improve utilization of energy in the Austrian region of Carinthia and the Italian region of Friuli-Venezia Giulia.

\subsection{Energy usage scenarios}
We initially carried out a survey study to highlight differences in terms of consumption scenarios, which affect the way inhabitants use energy and consequently the overall energy profile.
An outcome was the identification of electrical devices, as well as the penetration of generation from renewable energy \cite{monacchi:2013Nov}.
We found out that the use of electrical devices for cooking and heating purposes (i.e., electric boilers, heaters, hobs and ovens) is more diffuse in Carinthia than Friuli, where a more developed gas network can reduce electricity costs. 
Use of renewable energy is still limited, with photovoltaic systems having the highest penetration (7.91\% in FVG and 2.69\% in Carinthia).
Also, residents of Friuli tend to use air conditioners (45.19\% compared to the 2.16\% of Austrian respondents) and can already exploit a time-of-use pricing scheme due to the availability of automatic digital meters.
Householders in Friuli declared to already exploit favourable pricing conditions to operate their washing machine (62.59\%), lights (24.46\%), iron (22.3\%), electric oven (21.58\%), dryer (10.79\%), conditioner (10.07\%), and dishwasher (9.35\%).

The study also allowed for an estimation of energy usage in residential settings, as well as to gain insights on the inhabitants' attitute towards demand response and energy management systems \cite{tamer:smarthome}.
Inhabitants from Carinthia expressed the willingness to exploit multi-tariff pricing schemes, expecially for operating their washing machine (48\%), electrical boiler (23\%) and dryer (20\%).
Consumption reduction in the last 4 years was pursued by replacing appliances with more efficient ones, as done by 67.20\% in Carinthia.
\subsection{Measurement campaign}
The following step was to carry out a measurement campaign in selected households to investigate actual energy utilization.
In particular, we have been monitoring the following scenarios:
\begin{enumerate}
  \item A detached house with 2 floors in Spittal an der Drau (AT).
  The residents are a retired couple, spending most of time at home.
  \item An apartment with 1 floor in Klagenfurt (AT).
  The residents are a young couple, spending most of daylight time at work during weekdays, mostly being at home in evenings and weekend.
  \item A detached house with 2 floors in Spittal an der Drau (AT).
  The residents are a mature couple (1 housewife and 1 employed) and an employed adult son (28 years).
  \item A detached house with 2 floors in Klagenfurt (AT).
  The residents are a mature couple (1 working part-time and 1 full time), living with two young kids.
  \item An apartment with 2 floors in Udine (IT).
  The residents are a young couple, spending most of daylight time at work during weekdays, although being at home in evenings and weekend.
  \item A detached house with 2 floors in Colloredo di Prato (IT).
  The residents are a mature couple (1 housewife and 1 employed) and an employed adult son (30 years).
  \item A terraced house with 3 floors in Udine, (IT).
  The residents are a mature couple (1 working part-time and 1 full time), living with two young children.
  \item A detached house with 2 floors in Basiliano (IT).
  The residents are a mature couple, with one being retired and therefore spending most of time at home.
\end{enumerate}
\begin{table}[!htbp]
 \centering
 \caption{Monitored devices}\label{tab:devs}
 \rowcolors{1}{}{lightgray}
    \begin{tabular}{| c  m{0.8\columnwidth} |}
    \hline
    \textbf{House}		&	\textbf{Devices}\\
    \hline
    	0		&	Coffee machine, washing machine, radio, water kettle, fridge w/ freezer, dishwasher, kitchen lamp, TV, vacuum cleaner\\ 
		1		&	Fridge, dishwasher, microwave, water kettle, washing machine, radio w/ amplifier, dryier, kitchenware (mixer and fruit juicer), bedside light\\	
		2		&	TV, NAS, washing machine, drier, dishwasher, notebook, kitchenware, coffee	machine, bread machine\\ 
		3		&	Entrance outlet, Dishwasher, water kettle, fridge w/o freezer, washing machine, hairdrier, computer, coffee machine, TV\\ 
		4		&	Total outlets, total lights, kitchen TV, living room TV, fridge w/ freezer, electric oven, computer w/ scanner and printer, washing machine, hood\\ 
		5		&	Plasma TV, lamp, toaster, stove, iron, computer w/ scanner and printer, LCD TV, washing machine, fridge w/ freezer\\ 
		6		&	Total ground and first floor (including lights and outlets, with whitegoods, air conditioner and TV), total garden and shelter, total third floor. \\
		7		&	TV w/ decoder, electric oven, dishwasher, hood, fridge w/ freezer, kitchen TV, ADSL modem, freezer, laptop w/ scanner and printer\\ 
    \hline
    \end{tabular}
\end{table}
The main outcome was the GREEND dataset, containing power profiles of selected devices (see Table~\ref{tab:devs}) at 1 Hz resolution for a time span of about 1 year \cite{greend}.
\section{Room for intervention}\label{sec:analysis}
\subsection{Identifying energy hogs}
\begin{figure*}
        \centering
        \begin{subfigure}[b]{0.48\textwidth}
                \includegraphics[width = \columnwidth]{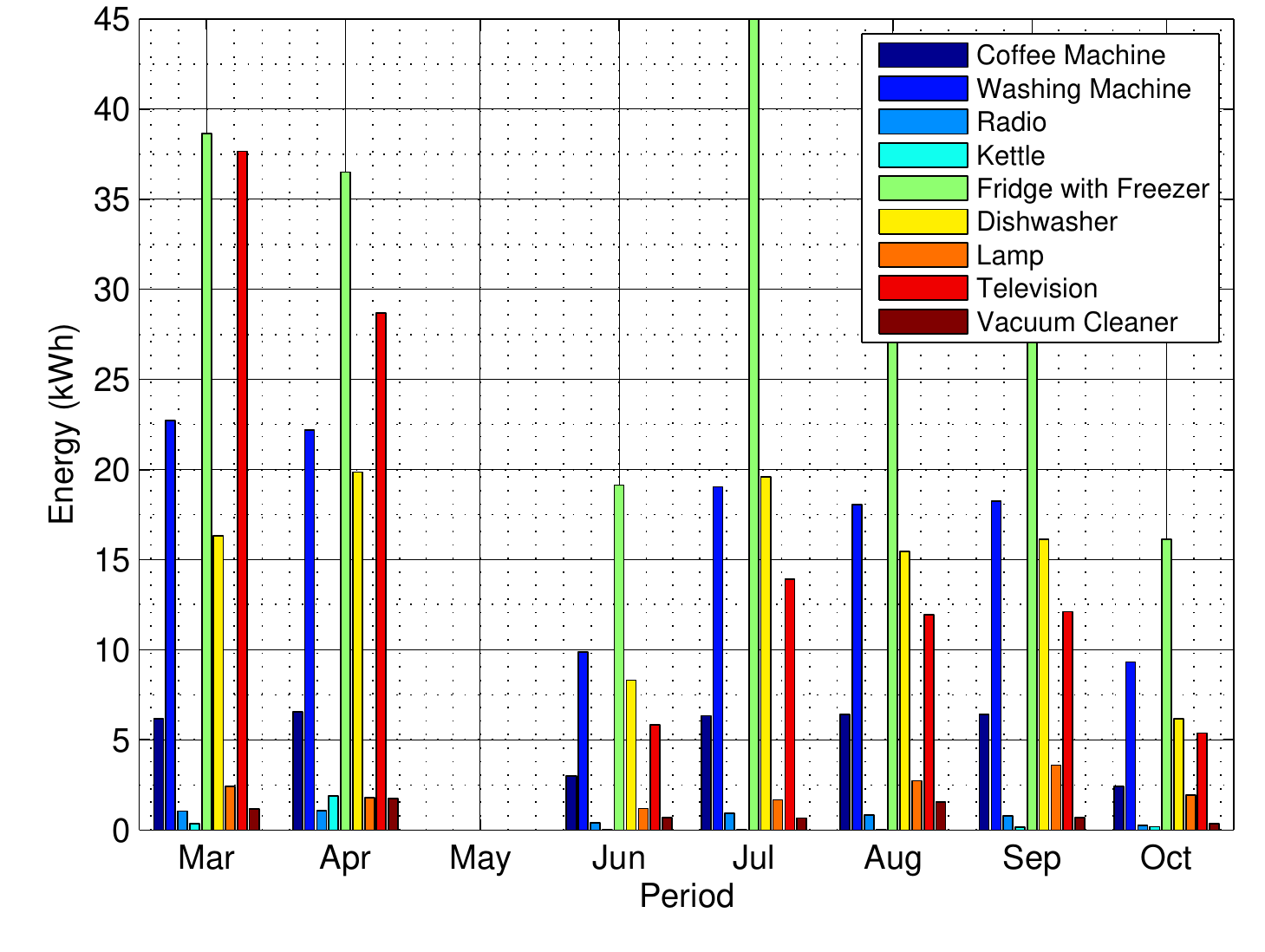}
                \caption{Power consumption of site S0}\label{fig:CAR:site0}
        \end{subfigure}%
        ~
        \begin{subfigure}[b]{0.48\textwidth}
                \includegraphics[width = \columnwidth]{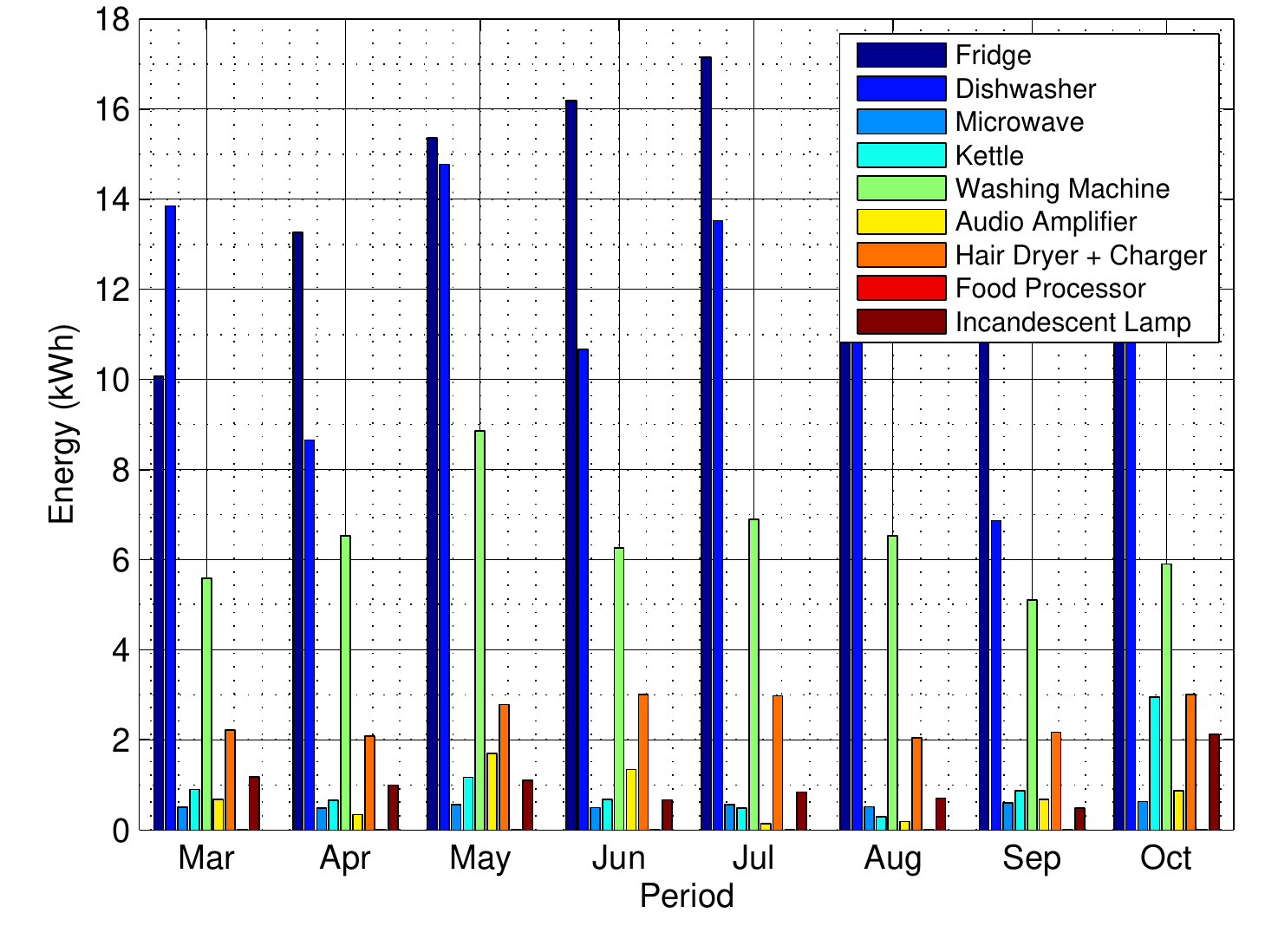}
                \caption{Power consumption of site S1}\label{fig:CAR:site1}
        \end{subfigure}
        ~ 
        \begin{subfigure}[b]{0.48\textwidth}
                \includegraphics[width = \columnwidth]{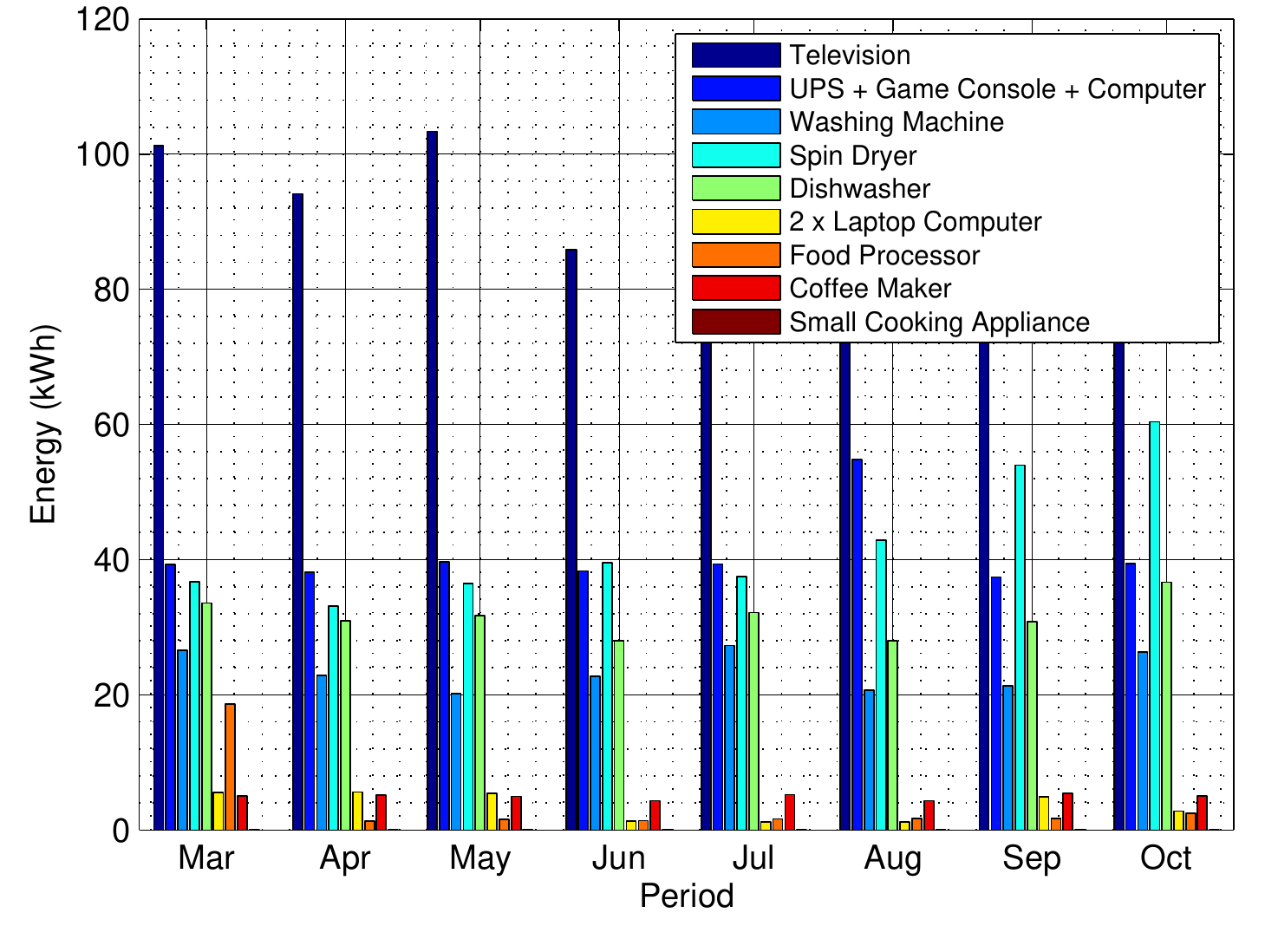}
                \caption{Power consumption of site S2}\label{fig:CAR:site2}
        \end{subfigure}
        ~
        \begin{subfigure}[b]{0.48\textwidth}
                \includegraphics[width = \columnwidth]{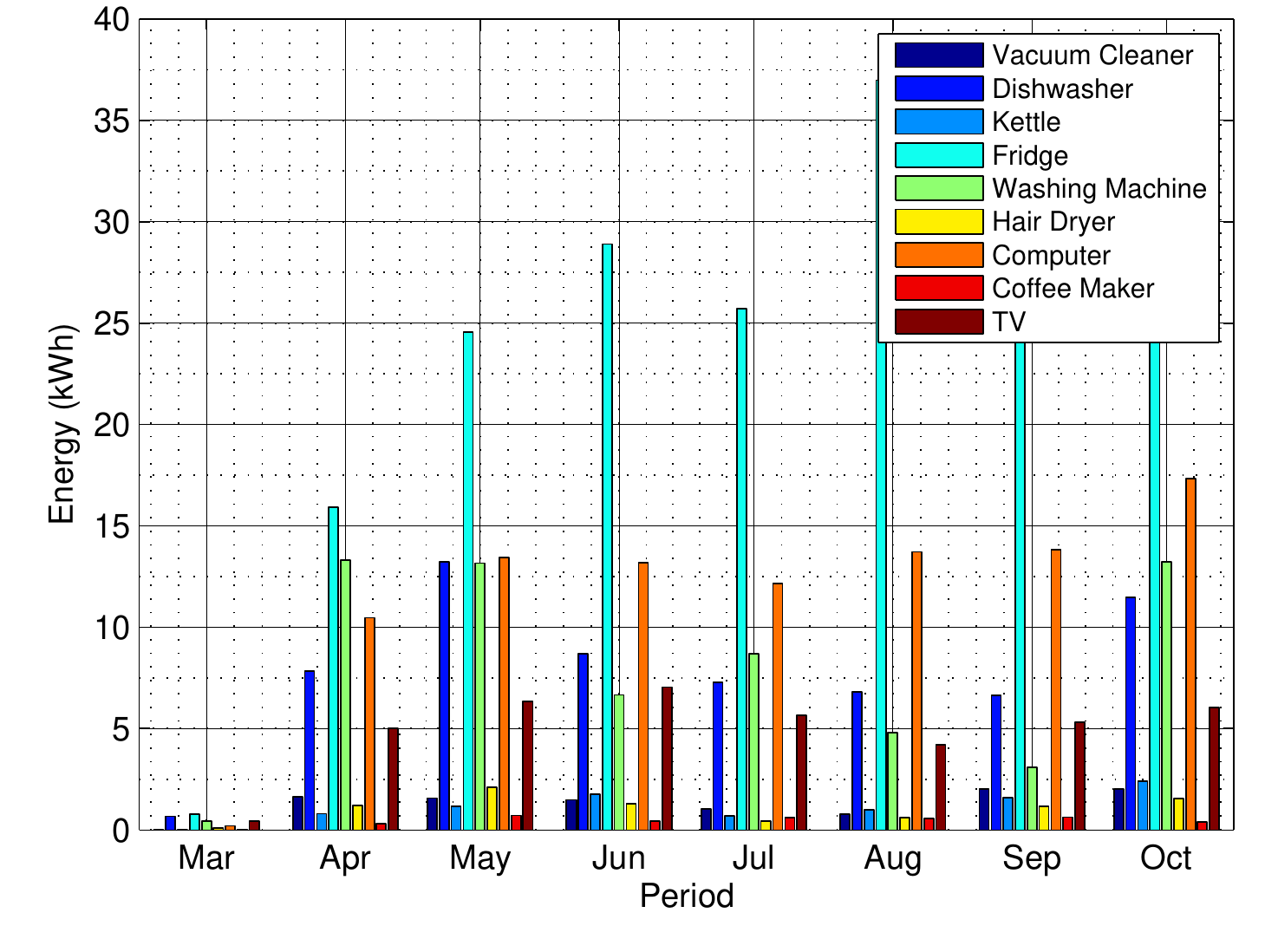}
                \caption{Power consumption of site S3}\label{fig:CAR:site3}
        \end{subfigure}
        \caption{Power consumption of monitored Austrian sites}\label{fig:power_consumption_CAR}
\end{figure*}

\begin{figure*}
        \centering
        \begin{subfigure}[b]{0.5\textwidth}
                \includegraphics[width = \columnwidth]{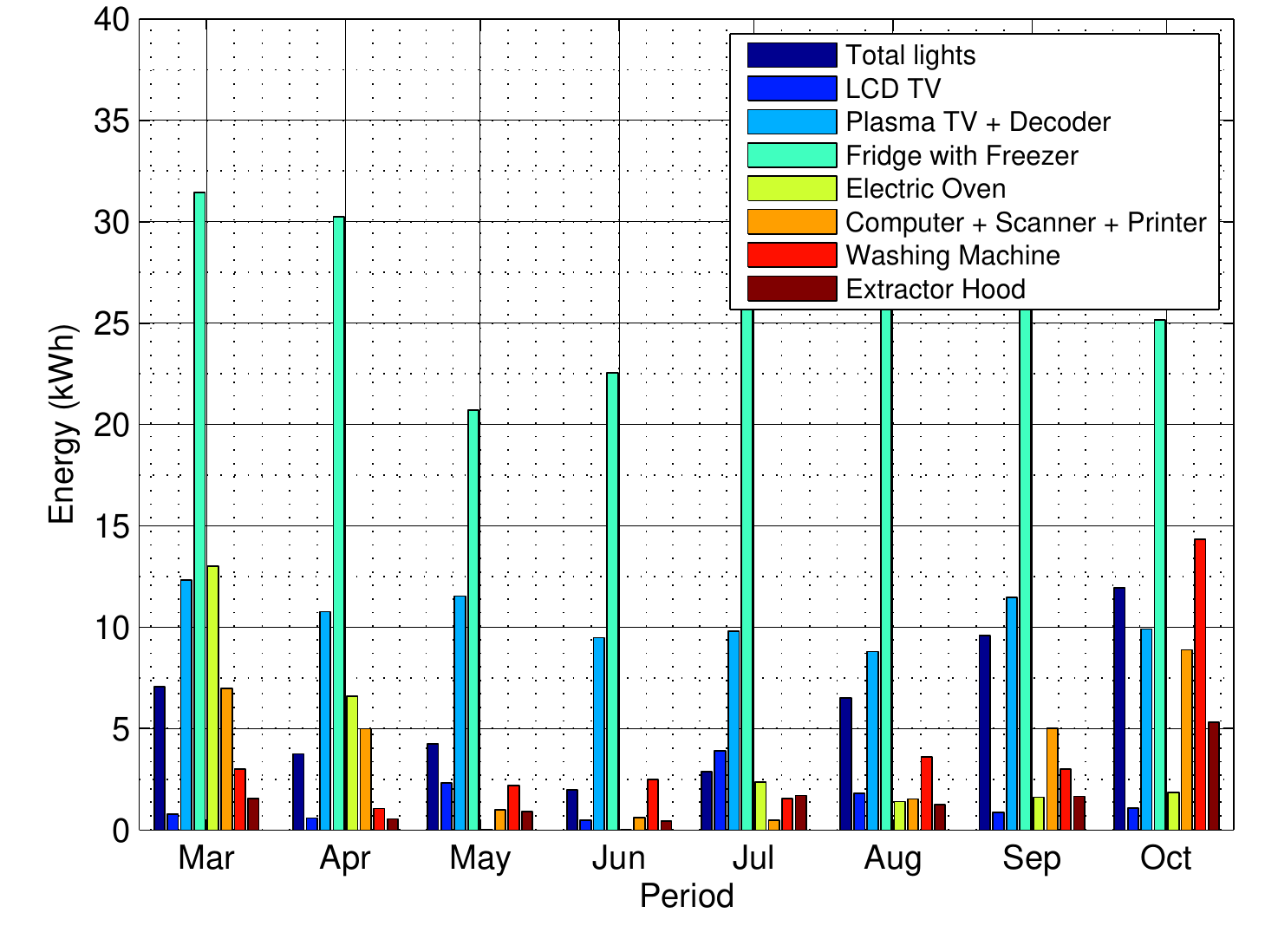}
                \caption{Power consumption of site S4}\label{fig:CAR:site4}
        \end{subfigure}%
        ~
        \begin{subfigure}[b]{0.5\textwidth}
                \includegraphics[width = \columnwidth]{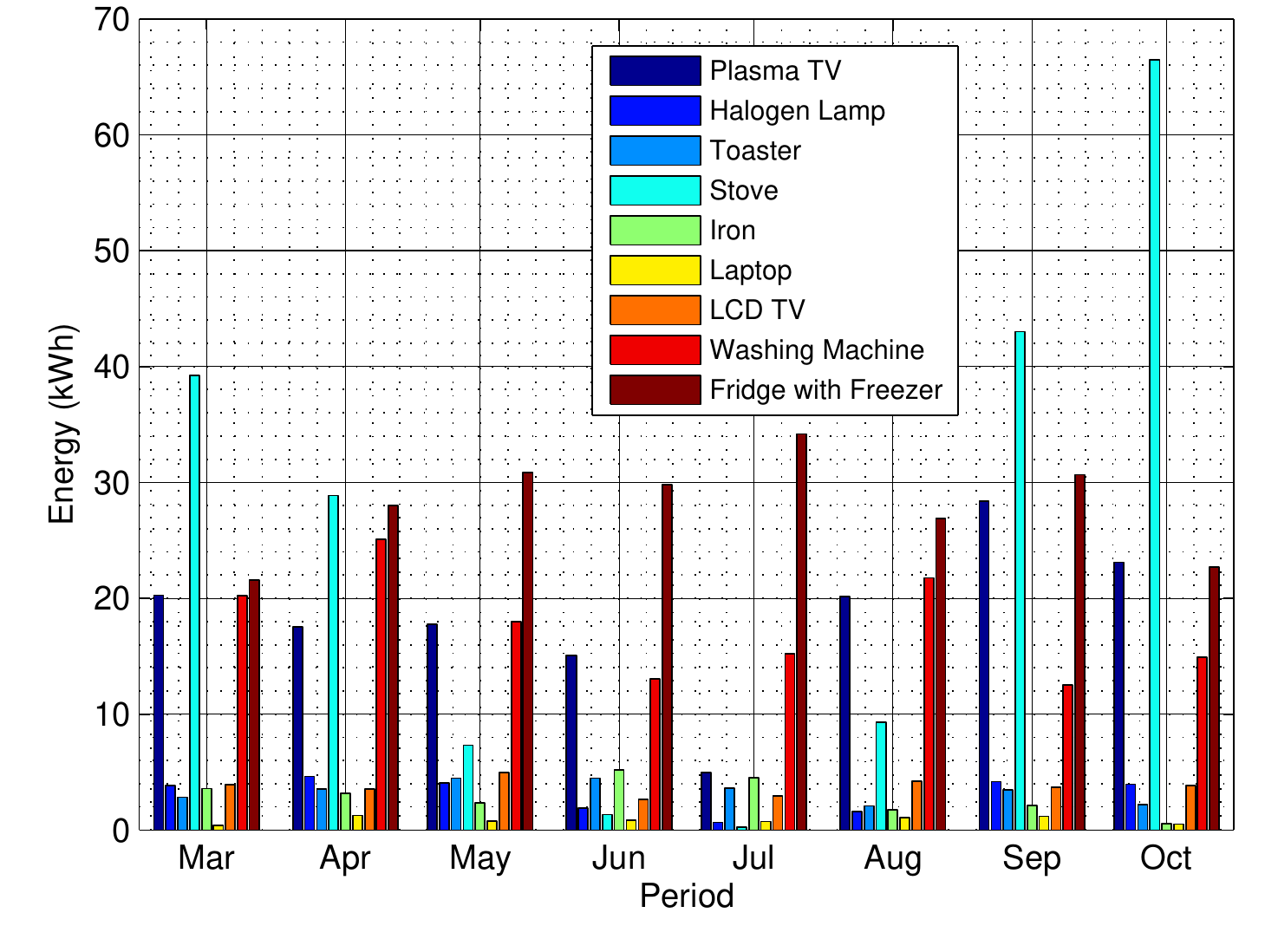}
                \caption{Power consumption of site S5}\label{fig:CAR:site5}
        \end{subfigure}
        
        \begin{subfigure}[b]{0.5\textwidth}
                \includegraphics[width = \columnwidth]{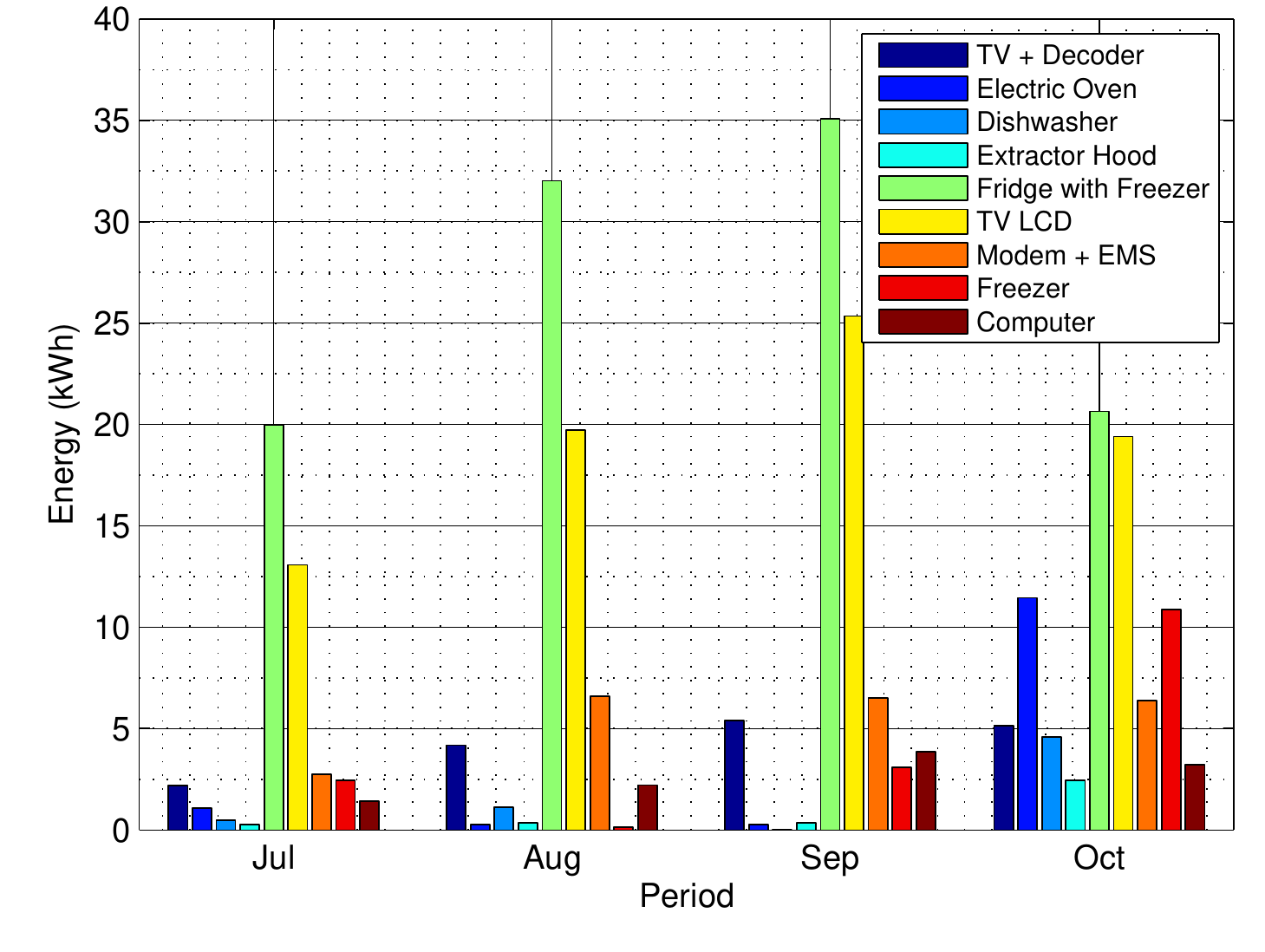}
                \caption{Power consumption of site S7}\label{fig:CAR:site7}
        \end{subfigure}
        \caption{Power consumption of monitored Italian sites}\label{fig:power_consumption_FVG}
\end{figure*}

This section provides an analysis of energy consumption in the monitored sites.
In particular, Fig.~\ref{fig:power_consumption_CAR} and Fig.~\ref{fig:power_consumption_FVG} show the energy usage respectively in the Austrian and Italian households.
We exclude from this discussion the 6th site shown in the previous section, as in such deployment we collected circuit-level measurements rather than individual devices.
As visible, the fridge is the most consuming device in all settings, determining between the 40\% and the the 47\% of monitored consumption.
Furthermore, consumption peaks are observable in the summer period due to higher environment temperature in the northern emisphere.
A considerable share is also accounted by the dryer, the dishwasher and the washing machine.
We also remark the presence of multiple incandescent lightbulbs in site S1, where for example, the bedside lamp alone determines the 2\% of monitored consumption. 
In site S2, a considerable contribution is given by the plasma TV, as well as the stand-by consumption of multiple consumer electronics devices (i.e., uninterruptible power supply, network attached storage, game console, personal computers).
Site S3 presents a similar situation, with the desktop computer accounting for the 22\% of monitored consumption, which translates into more than 12 kWh every month.
As for the Italian deployments, the situation reflects the one previously presented.
The fridge is the device responsible for the largest consumption, accounting for between 24\% and 46\% of the total monitored consumption.
Televisions have also a considerable impact, being responsible for 20\%, 25\% and 39\% of total consumption in S4, S5 and S7, respectively.
To understand the operational costs of televisions, it is interesting to remark that in site S4 and S5 the TVs consume more than the washing machine, a device commonly considered as energy-hungry.
Accordingly, in site S4 and S5 the washing machine is responsible for only 5\% and 10\% of the total monitored consumption in September.
In site S5, the consumption of the washing machine and the electric iron is lower than that of televisions.
The analysis includes the consumption of both plasma and LCD televisions, although the former type is generally demanding higher power.
This is due to technical and cost reasons. 
Due to they higher contrast ratio and larger viewing angle, plasma televisions tend to be installed in the day area of the household, 
such as the living room, which implies longer usage of the device. This suggests possible improvements, as discussed in Sect.~\ref{sec:shifting}.
\subsection{Exploitability of time-dependent energy tariffs}
Time-dependent energy tariffs are meant to foster postponement of device operation to off-peak periods.
As previously analyzed in \cite{monacchi:2013Nov}, in Italy more than 32 millions digital meters were installed by ENEL, the main italian distribution system operator (DSO).
The advanced metering functionalities provided by the digital meters allow for automatic meter reading and enables more dynamic energy pricing.
This differs from the situation in Austria, especially in the Land of Carinthia, where the roll out of digital meters at large scale is yet to be started.  
Consequently, exploitability of time-dependent energy tariff is not yet possible in Austria.
 
\begin{figure*}[h!]
	\centering
	\begin{subfigure}[b]{0.48\textwidth}
		\includegraphics[width = \columnwidth]{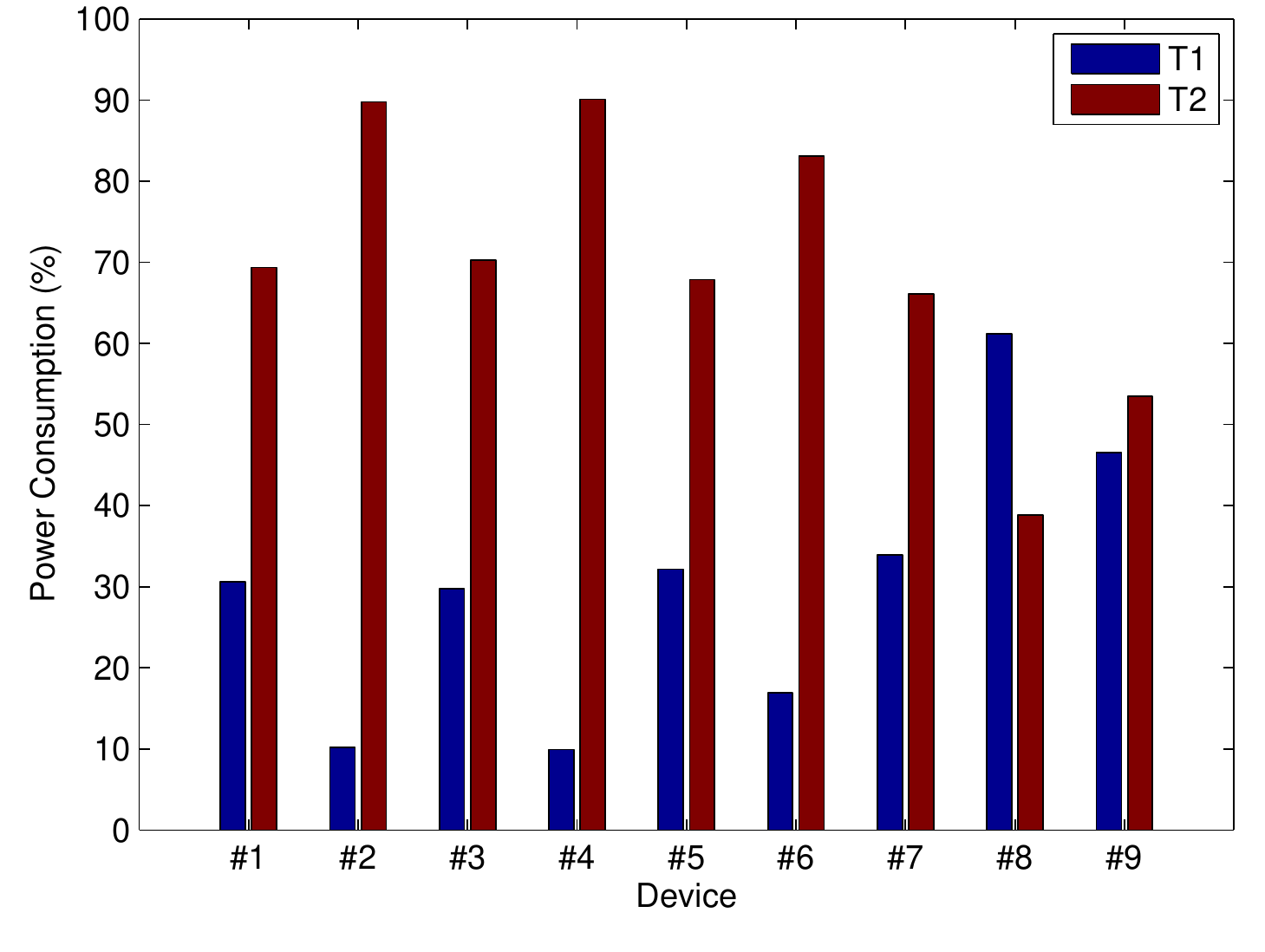}
		\caption{\label{fig:bills:site4}Energy consumption per device and time slots of site S4.}
	\end{subfigure}
	~
	\begin{subfigure}[b]{0.48\textwidth}
		\includegraphics[width = \columnwidth]{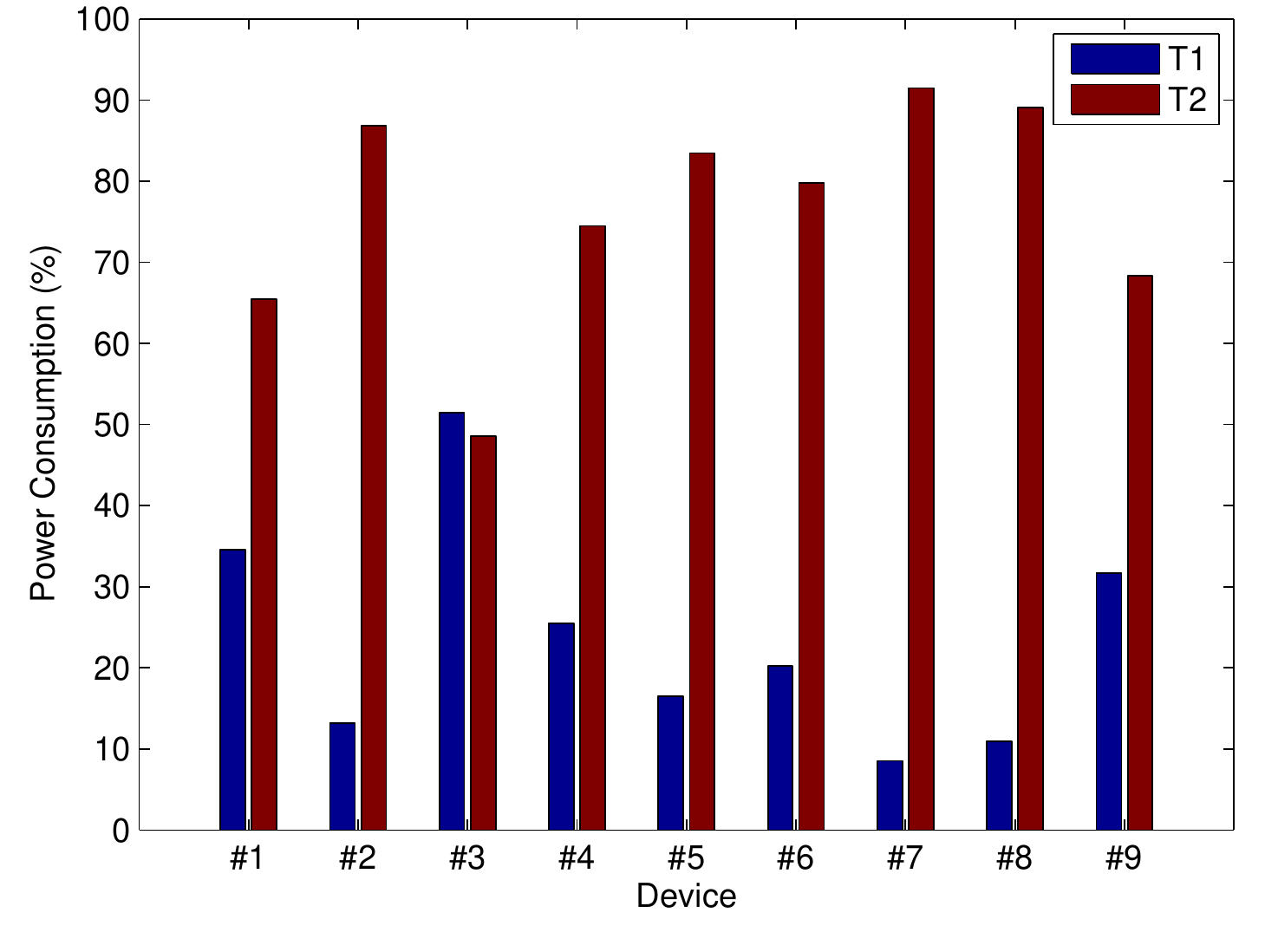}
		\caption{\label{fig:bills:site5}Energy consumption per device and time slots of site S5.}
	\end{subfigure}
	~
	\begin{subfigure}[b]{0.48\textwidth}
		\includegraphics[width = \columnwidth]{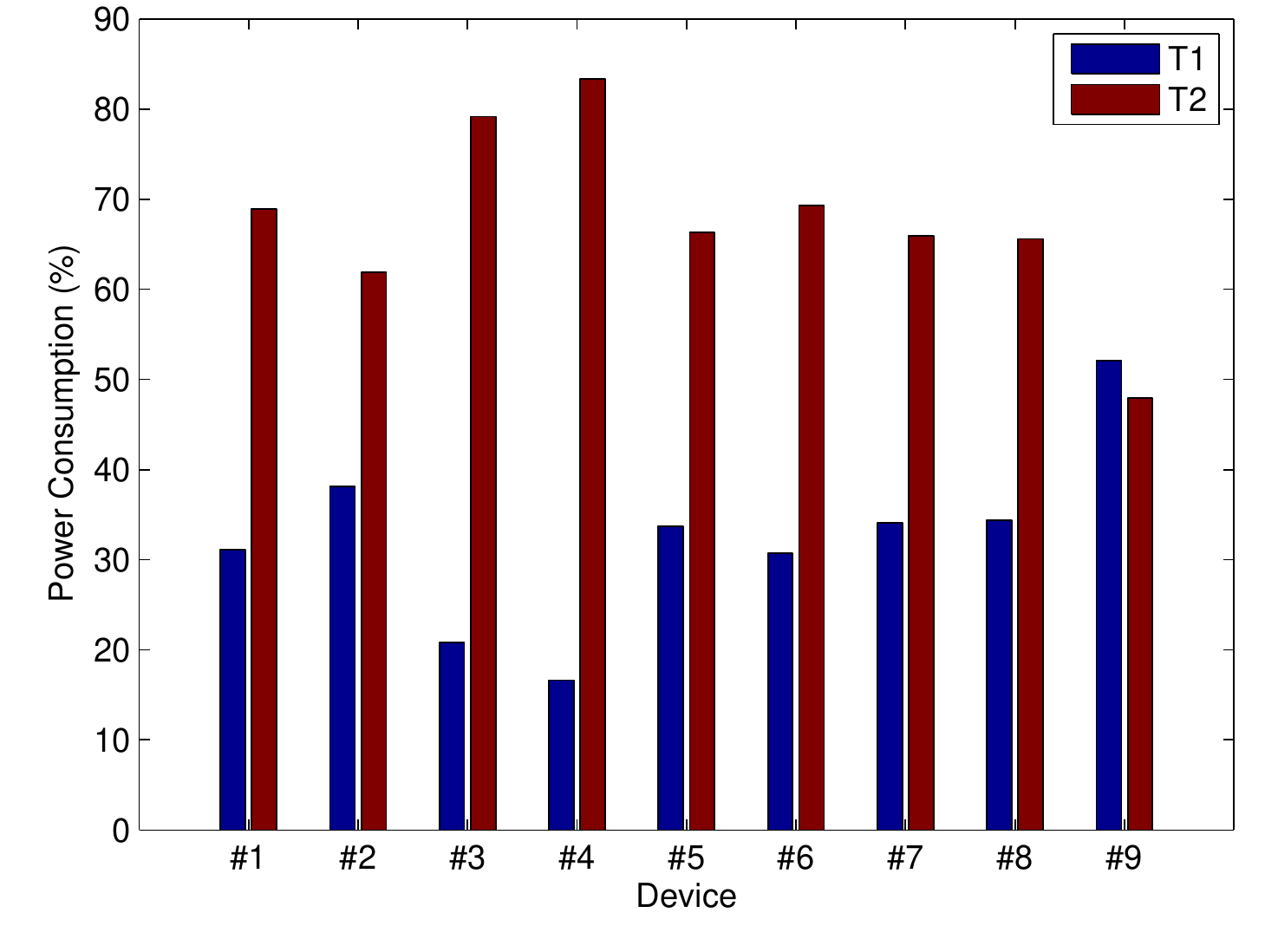}
		\caption{\label{fig:bills:site7}Energy consumption per device and time slots of site S7.}
	\end{subfigure}
	\caption{Energy consumption per device and time slots of Italian sites}
\end{figure*}

Let us try to quantify energy costs and potential savings in Italy by considering the offer of ENEL for residential users (contracts below 3kW), 
which is divided in two slots, namely:
\begin{itemize}
  \item T1, from Monday to Friday between 8 AM and 7 PM
  \item T2, for the night hours and weekends, as well as during public holidays
\end{itemize}
To promote efficiency, in each time slot the price also varies depending on four categories of consumption (see Table \ref{tab:bills:1}).
\begin{table}[h]
	\centering
	\caption{\label{tab:bills:1} Energy consumption categories.}
	\begin{tabular}{c | c c }
	 & Lower Bound & Upper Bound \\
	Category & (kWh/year) & (kWh/year) \\
	\hline 
	C1 & - & 1800 \\
	C2 & 1801 & 2640 \\
	C3 & 2641 & 4440 \\
	C4 & 4441 & - \\
	\end{tabular}	
\end{table}
Table \ref{tab:bills:2} shows the energy cost of each category and the entries contributing to such cost.
\begin{table}[h]
	\centering
	\caption{\label{tab:bills:2} Cost of energy in Italy.}
	\begin{tabular}{ c c  | c | c | c | c }
	 & & Energy & Delivery Price & Grid & \\
	 & & Price & Fixed + Variable & Services & TOT \\
	 & & (Eur/kWh) & (Eur/kWh) & (Eur/kWh) & (Eur/kWh) \\
	 \hline
	 \multirow{4}*{T1} & C1 & 0.067310 & 0.01381&  0.046392 & 0.127512\\
    & C2 & 0.06731 & 0.01711& 0.101512& 0.185932 \\
	& C3 & 0.06731 & 0.02066 & 0.165762 & 0.253732 \\
	& C4 & 0.06731 & 0.02446 & 0.208432 & 0.300202 \\
	\hline
	 \multirow{4}*{T2} & C1 & 0.06094 & 0.01381 &  0.046392 & 0.121142 \\
    & C2 & 0.06094 & 0.01711 & 0.101512 & 0.179562 \\
	& C3 & 0.06094 & 0.02066 & 0.165762 & 0.247362 \\
	& C4 & 0.06094 & 0.02446 & 0.208432 & 0.293832 \\
		\end{tabular}	
\end{table}
Concerning the monitored sites, they denote a rather representative case, as they span from the lowest C1 to the highest C4 energy consumption category (see Table~\ref{tab:bills:3}).
\begin{table}[h]
	\centering
	\caption{\label{tab:bills:3}Per-year  consumption and price category of Italian sites.}
	\begin{tabular}{c | c | c }
	& \multirow{2}*{Category} & Energy Consumption\\
	& & (kWh/year) \\
	\hline
	Site 4 & C1 & 1277\\
	Site 5 & C4 & 4778 \\
	Site 6 & C3 & 3349 \\
	Site 7 & C3 & 4099 	
	\end{tabular}
\end{table}
Let us now investigate the distribution of device usage over the time slots.
Figs.~\ref{fig:bills:site4}, \ref{fig:bills:site5} and \ref{fig:bills:site7} show energy consumption of each monitored device, indicated in percentage terms so that the sum of power for each time slots is 100\%. 
As visible, users are aware of the incentives provided by the time-slotted tariff, being the monitored devices operated mostly during T2.
Users in site S5 better exploit than others the tariff T2, given the larger consumption spread between T1 and T2, especially for the washing machine (\#8) and the iron (\#5).
Clearly, not all devices can be scheduled to a different time slot, such as the toster (\#3).
While users' awareness of time-slotted tariffs is rather high in Italy \cite{monacchi:2013Nov}, the actual savings provided by operating in T2 is minimal.
For instance, in S4 the washing machine is being used for 19 kWh in T1 and 12 kWh in T2, which accounts for 3.9 Eur.
Shifting completely the operation to T2 would yield savings for about 12 cents, which is not enough to foster a behavior change.
\subsection{Energy efficiency policies}\label{sec:policies}
From the analysis, multiple ways of improving energy efficiency emerged, including:
\begin{enumerate}
  \item \textbf{lighting} promoting replacement of incandescent bulbs with energy saving ones;
  \item \textbf{device diagnostics} promoting replacement of old appliances with more energy efficient ones, expecially regarding white goods but also involving consumer electronics (e.g., LCD/LED TV in place of a plasma TV);
  \item \textbf{shedding of standby losses} promoting switching-off of consumer-electronic devices when people are not likely to be at home, such as ADSL modems and TVs;
  \item \textbf{device shifting} promoting postponement of particularly energy demanding devices to off-peak periods, in order to operate loads in cheaper time periods. 
   This includes both deferral and preference of efficient devices to energy demanding ones, as it will be shown in Sect.~\ref{sec:shifting}.
\end{enumerate}
While these policies have a general validity, the benefits of the data analysis is limited to the users involved in the campaign.  
In order to extend the analysis to a large area, we propose in the next section an open solution to provide automatic energy advice.
\section{System design}\label{sec:platform_solution} 
\begin{figure}[h]
	\centering
	\includegraphics[width=0.7\columnwidth]{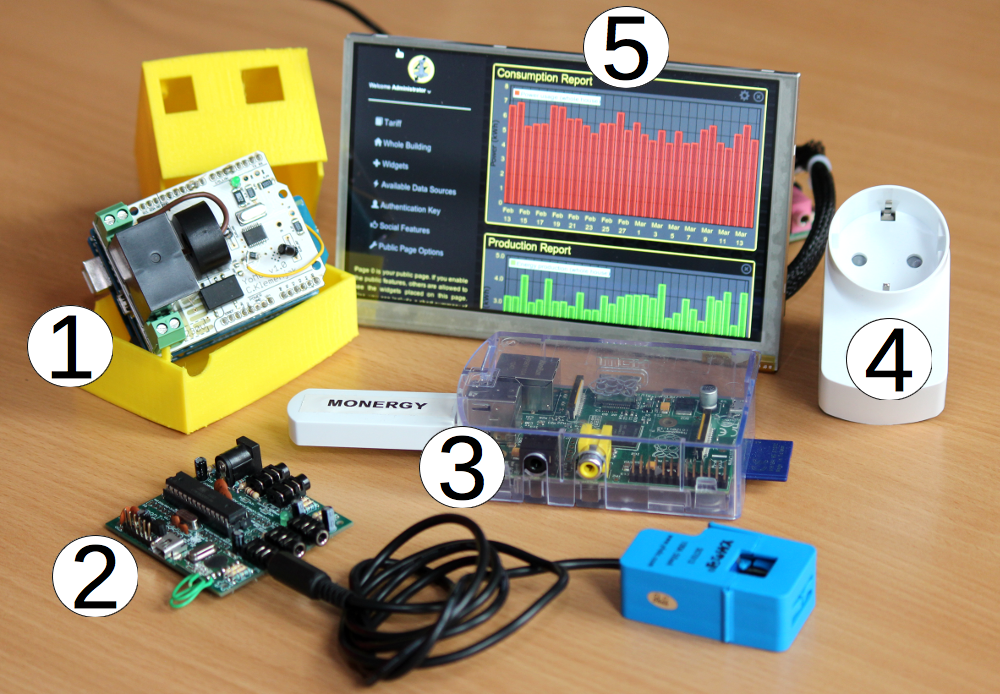}
	\caption{The measurement hardware and the system interface}\label{fig:system_picture}
\end{figure}

The analysis demonstrated the need for more effective energy management strategies, which consider both energy usage behavior and production from renewables.
To this end, we propose a system consisting of (see Fig.~\ref{img:archi}):
\begin{figure*}[tH!]\centering
	\includegraphics[trim=2.5cm 25cm 4cm 2cm,clip,width=0.9\textwidth]{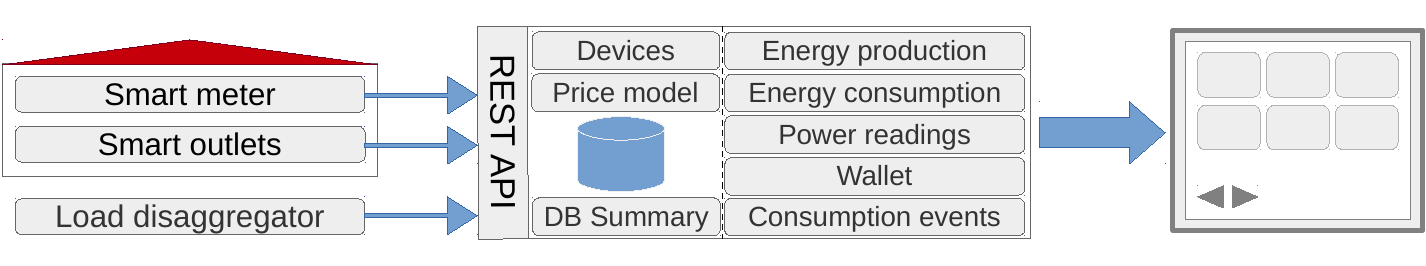}
\caption{The system architecture}\label{img:archi}\end{figure*} 
\begin{description}
  \item[\textbf{an aggregate power meter.}]
 	For a basic analysis the meter should measure active power (W), while apparent (VA) and reactive power (VAR) can be useful for more advanced analyzes such as load disaggregation.
 	Consequently, aggregated production and consumption data of the dwelling can be computed by averaging multiple power readings over time.
 	Such a meter needs to provide the required measurement with sufficient frequency (for example 1~Hz), although unlike meters used for billing purposes measurement accuracy is relaxed.
 	The YOMO open-hardware meter previously designed and proposed in \cite{yomo} offers a low-cost metering solution for smart meters and smart plugs.
 	The WiTiKee power meter\footnote{http://www.witikee.com} is a commercial solution that provides high accuracy and overcomes the network coverage limitations identified in \cite{energycon}, by exploiting a combined powerline-wireless network. 
  \item[\textbf{an appliance-level monitoring system.}] 
  	To collect appliance-level data a network of sensing units can be used, such as the OpenEnergyMonitor\footnote{http://openenergymonitor.org/emon/} and the commercial Plugwise\footnote{https://www.plugwise.com}. 
  	Events can be locally detected using edge-detection techniques (e.g., thresholding), as well as extracted from aggregate power readings using a third-party non-intrusive load monitoring tool~\cite{nilm,Egarter2014,egarter}. 
  	To this end, we provide a Python script capable of detecting events from power measurements being collected at device-level\footnote{https://sourceforge.net/projects/monergy/files/Gateway}.
  	In absence of an event detection stage, events can also be manually defined by users after they occured \cite{Costanza:2012}.
  \item[\textbf{an energy management system}]
  	which includes a web-based analysis tool and a mobile application.
  	A REST interface~\cite{Fielding:2000} was implemented to provide accessibility to processed data to multiple applications and easily integrate multiple data sources. 
	To ensure privacy each method call is authenticated through a private token. 
	Moreover, to connect the analysis tool and the mobile application the Google cloud messaging (GCM)\footnote{https://developer.android.com/google/gcm/index.html} infrastructure is used, in order to send push messages.
\end{description} 
%
Fig.~\ref{fig:system_picture} shows the measurement hardware and the system interface. Numbered items from left are i) the YOMO, ii) the OpenEnergyMonitor, iii) a Raspberry Pi acting as home gateway, iv) a Plugwise smart plug and v) a Udoo board\footnote{http://www.udoo.org} with a touch screen acting as in-home display.
\section{Mjölnir: an open-source energy advisor}\label{sec:advisor}
Following the requirements identified in Sect.~\ref{sec:introduction}, we developed a web-based energy management system capable of analyzing energy consumption and production data, from both aggregate and disaggregated sources.
\begin{figure}[h!]
	\centering
	\includegraphics[width=0.7\columnwidth]{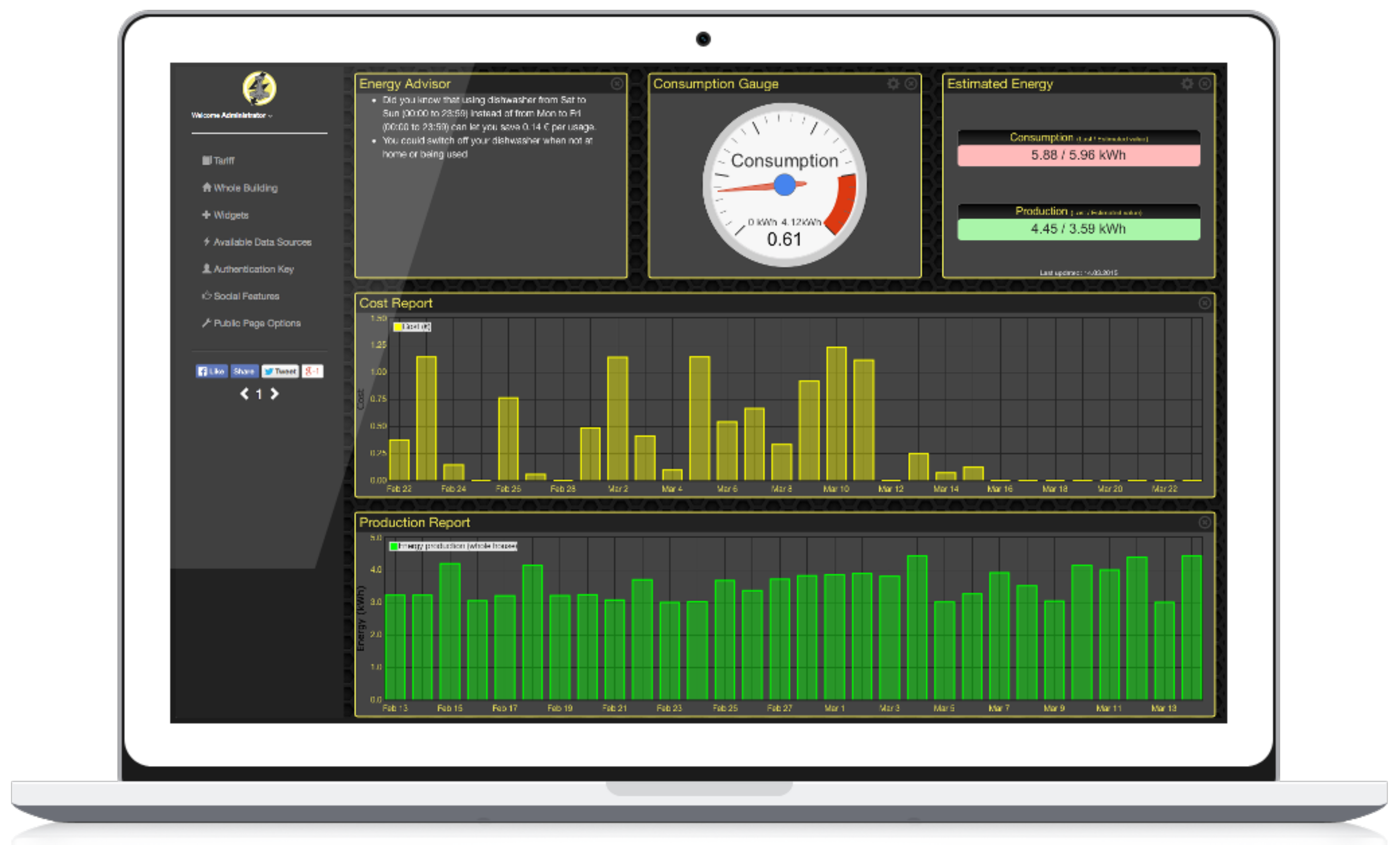}
	\caption{The Mjölnir interface 0.2}\label{fig:screenshot}
\end{figure}
This includes: i) aggregated building production and consumption information, ii) power readings from selected circuits and devices and iii) energy usage events espressed as tuple $(device,$ $t_{start},$ $duration,$ $E_{consumption})$.
The framework, named Mjölnir, is released as open-source and available as SourceForge project\footnote{http://mjoelnir.sourceforge.net}.
Mjölnir is implemented in PHP~5 and uses MySQL as DBMS. The front-end dashboard is implemented in HTML~5, CSS~3 and Javascript.
At the version 0.2 Mjölnir provides:
\begin{description}
  \item[ \textbf{credit-based device management} ]
  	where each monitored device is associated to a credit, which is decreased upon device usage \cite{Monacchi:2013:insert_coin}.
  	This provides a fine-grained consumption resolution, providing an understanding of the cost of operating each device.
  \item[ \textbf{device metadata} ] to annotate each device.
  	To determine the applicability of certain analyses, each device is described through the following attributes: type (e.g., fridge), mobility and room, curtailability, autonomy (i.e., user control) and stand-by mode.
  	In particular, the room and device metadata are based on the large vocabulary introduced in \cite{NILM_Metadata}. This makes the integration with analysis frameworks sharing this data model (e.g., nilm toolkit~\cite{Batra:2014}) possible.
  \item[ \textbf{tariff-based energy data analysis} ]
  	The analysis tool is based on a price model, which is expressed as energy tariffs.
  \item[ \textbf{modular interface} ] 
  The interface exploits Twitter's Bootstrap\footnote{http://getbootstrap.com} libary in order to be seamlessy visualised on both mobile terminals (e.g., smartphones and tablets) and computers.
  Moreover, the framework is organized on pages and cells and based on the concept of widget, which provides both modularity and flexibility to the interface structure.
  A widget can provide various features, such as displaying charts or forecasting energy consumption, and can be placed in those cells.
  This also allows users to progressively adapt the feedback system in order to display energy information in a language that is meaningful to them: by placing things next to each other and only concentrate on interesting matters.
  \item[ \textbf{social features and public profile page} ]
  	Social features allow users for sharing their performance with their peers, including social networks and blogs.
  	To this end, we distinguish public and private pages of users. A public page represents the user's public profile.
  	Also, actual and estimated energy consumption and production for the current day is provided as a summary that can be embedded in external web pages (e.g., blogs). 
\end{description}
Currently available widgets are:
\begin{description}
  \item[ \textbf{production and consumption report} ] showing daily energy information over the last month;
  \item[ \textbf{cost report} ] showing daily energy cost over the last month;
  \item[ \textbf{production and consumption gauges} ] showing energy use for the current day;
  \item[ \textbf{energy estimation} ] showing an estimation of energy production and consumption for the current day, as based on the previous days;
  \item[ \textbf{device itemization} ] showing the consumption and cost per device, over the current day, week and year;
  \item[ \textbf{timeline} ] showing energy usage events per device, and described by their consumption and cost;
  \item[ \textbf{energy advisor} ] returning messages based on usage behavior in order to increase efficiency;
  \item[ \textbf{appliance usage} ] showing the usage models of user-driven devices;
\end{description}
\section{Providing tailored energy efficiency advice}\label{sec:advices}
The main focus of this study is the implementation as automatic advices of the policies identified in Sect.~\ref{sec:policies}.
In a first stage, candidate advices are formulated.
An advisor widget displays the list of advices and allows the user for their acceptance or rejection (See Fig.~\ref{fig:advisor_widget}).
We use the term conversion to indicate that the user explicitly accepts the recommendation, indicating the conversion of the advice into a behavior.
Once the conversion occurs the advice should not be recommended again, in order to minimize user's discomfort. 
An information filtering mechanism is implemented for the purpose.

\subsection{Formulating candidate advices}
Each candidate advice is formulated considering users' device usage events and targeted device type, as follows:
%
%
%
%
%
\begin{itemize}
  \item \textbf{Device diagnostics} advices replacement of appliances and it is thus useful to improve non-user-driven devices (e.g., fridge)
  	\begin{enumerate}
  	  \item Select non-user-driven devices
  	  \item Compute average consumption for each device type for all users\footnote{Can be done periodically and cached in a separate location}
  	  \item Retrieve devices whose average consumption is higher than the one for the device type of a certain threshold $\tau_{1}$ (e.g., 30\%) and suggest replacement 
  	\end{enumerate}
  \item \textbf{Device shifting}
  	\begin{enumerate}
  		\item Select user-driven devices
		\item Rank devices by their average consumption (according to consumption events)
		\item Rank tariffs by cost in order to select the best and worst tariffs available
		\item Suggest to use the device in the cheapest tariff and report the potential savings computed as $s = (l*t) - (l*c)$, respectively with $l$ average consumption for the device, $c$ and $t$ cheapest and most espensive energy tariffs.  	  
  	\end{enumerate}
  \item \textbf{Shedding of standby losses} suggests to switch-off devices in standby mode (such as displays, decoders, DVD players, battery chargers without load, air-conditioning systems) in periods of not use (e.g., night).
  This tip can be always returned for devices with a standby mode, although higher effectiveness can be achieved by exploiting an occupancy model.  
  \item \textbf{Device curtailment and moderate usage}
  	\begin{enumerate}
  	  \item Select user-driven devices;
  	  \item Rank devices by their positive deviation from the average number of usage for the device type and cost;
  	  \item Suggest to reduce the amount of times the device is being used and compute the yearly savings by multiplying the running cost spent for the current month;
  	\end{enumerate}
\end{itemize}
%
%
\subsection{Information filtering mechanism}
Once candidate advices are formulated, an information filtering mechanism is necessary.
This allows for presenting the most effective advices and limit the information overload, based on historical feedback.
We use the term conversion to indicate that the user explicitly accepts the recommendation, indicating the conversion of the advice into a behavior.
A feedback to an advice can be formalized through the tuple: $(user, advice\_type, device\_type, action, time)$.
In this way, it is possible to omit advices which were previously converted into a behavior (i.e., goal) or involving device types and advice types with low acceptance (i.e., negative feedback).

\begin{figure}[h!]
	\centering
	\shadowimage[width=0.6\columnwidth]{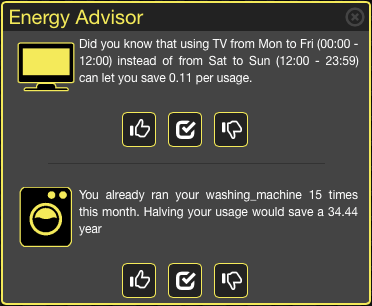}
	\caption{The advisor widget}\label{fig:advisor_widget}
\end{figure}

User feedback to recommendations can be explicit or implicit, depending on the possibility to directly espress acceptance or interest on the proposed items \cite{Jawaheer:2010}.
For explicit feedback this includes both rating scales and pursuing the suggested advice on the system by performing specific actions.
On the other hand, implicit feedback is built through inferences about the user's behavior, which makes the solution application dependent.
An example might be the frequent listening of an audio track to denote the user's interest.
Clearly, implicit feedback can only indicate positive feedback, as not listening to a track does not imply disliking it.
Consequently, explicit feedback offers a more complete and accurate picture of user's preferences.
This issue becomes clearer in our scenario, in which feedback is used to determine the degree of persistence of advices.
The results reported in \cite{persistence} showed that displaying multiple times the same recommendation does not improve the conversion rate unless the user has a big opinion drift. 
Consequently, we use the following 3-item Likert scale: ``Ok thanks'', ``I'm already doing it'', ``No thanks'' (See Fig.~\ref{fig:advisor_widget}).
This means that each advice can be formalized through the tuple $(user, advice\_type, device\_type, enabled, score)$.
A feedback of kind ``I'm already doing it'' indicates the right conversion of the advice into a behavior, which causes the deactivation of the advice.
A usefulness score is then computed for active advices using the votes resulting from ``Ok thanks'' and ``No thanks''.
Such value is used to rank the advices, while randomness is used to order advices with same usefulness value.
Positive feedback reinforces the advice by increasing its score, whereas negative feedback can result from a reluctance in operating the device or a mistrust in the specific advice type.    
Upon clicking on ``No thanks", the user is asked to select one of the two causes.
Based on this information, we decrease the score of all advices of the same type, that is, they either involve the same advice type or device type.
\section{Calculating the impact of intervention}\label{sec:evaluation} 
In this section we assess the proposed policies by providing an estimation of potential savings. 
In particular, based on the GREEND dataset we show that implementing the proposed policies can yield up to 34\% savings.
\subsection{Device diagnostics and replacement}
Replacing appliances with newer ones is a possible solution to increase efficiency.
Old appliances are inefficient due to technological progress and degradation from aging factors.
An example is given by two old fridges installed in site S5.
These devices do not belong to the set of monitored appliances available in GREEND due to coverage limitations of the selected monitoring platform.
For such reason, a dedicated measurement was carried out for a week on the site.
The measurements revealed an energy consumption of about 47.7 Wh and 28.6 Wh, for a total amount of 56 kWh per month and 668 kWh per year.
This number could be reduced to below 258 kWh per year by replacing the two freezers with two freezers belonging to A+++ energy class\footnote{The computation of energy consumption given by A+++ freezer has been made by assuming that the energy efficiency index (EEI) is equal to 22, the volume of the freezer is equal to 302 liters, and the appliance category is the 7.}.
The resulting energy saving would be equal to 34 kWh per month that corresponds to the 11\% of the total energy consumption of site S5. 
\begin{figure}[h!]
	\centering
	\includegraphics[width = 0.6\columnwidth]{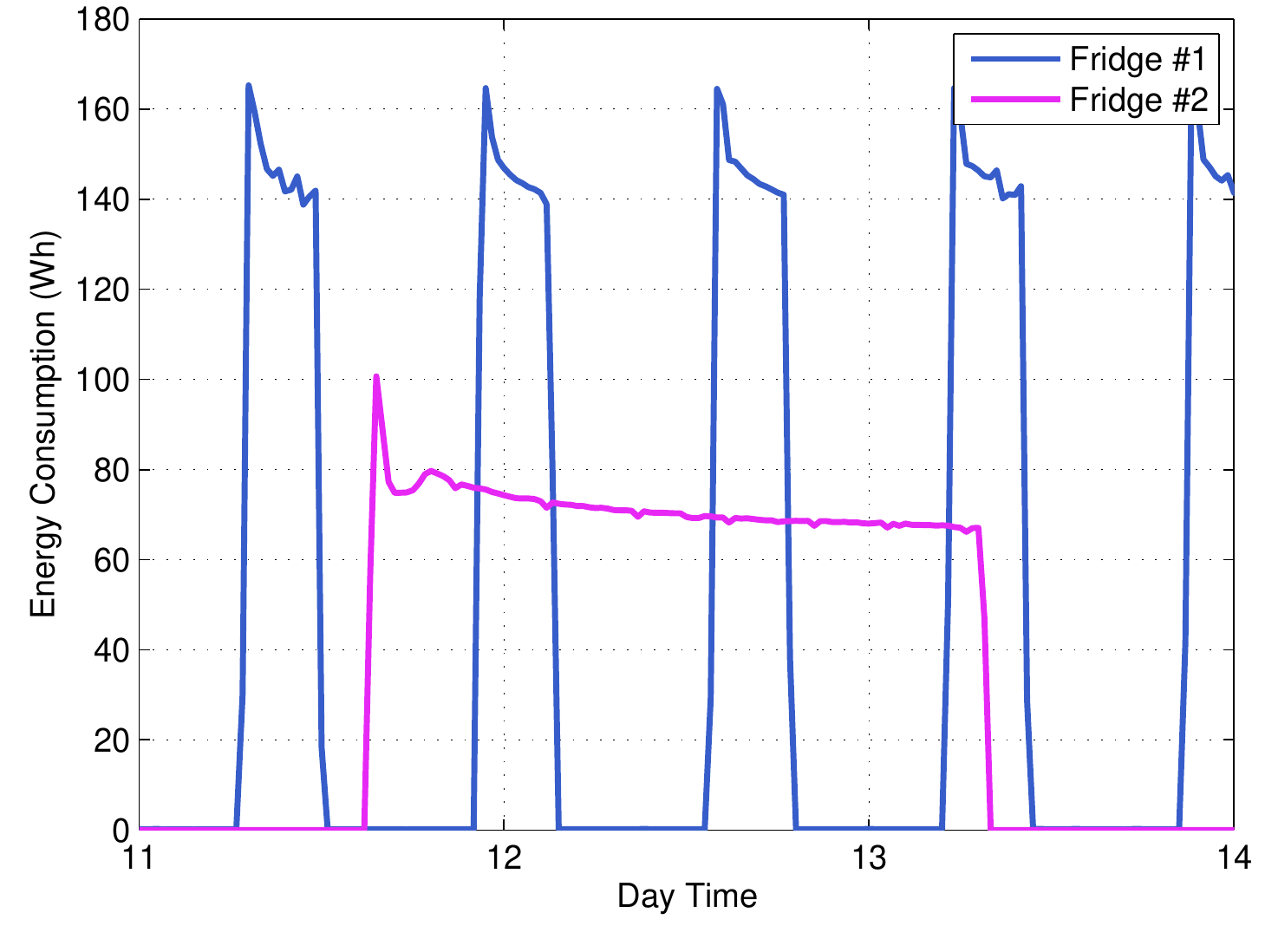}
	\caption{\label{fig:tip:3}Power consumption of an old fridge and an old freezer in site S5 during a time interval of three hours.}
\end{figure}
\subsection{Shedding of standby losses}
\begin{figure}[h!]
	\centering
	\includegraphics[width= 0.6\columnwidth]{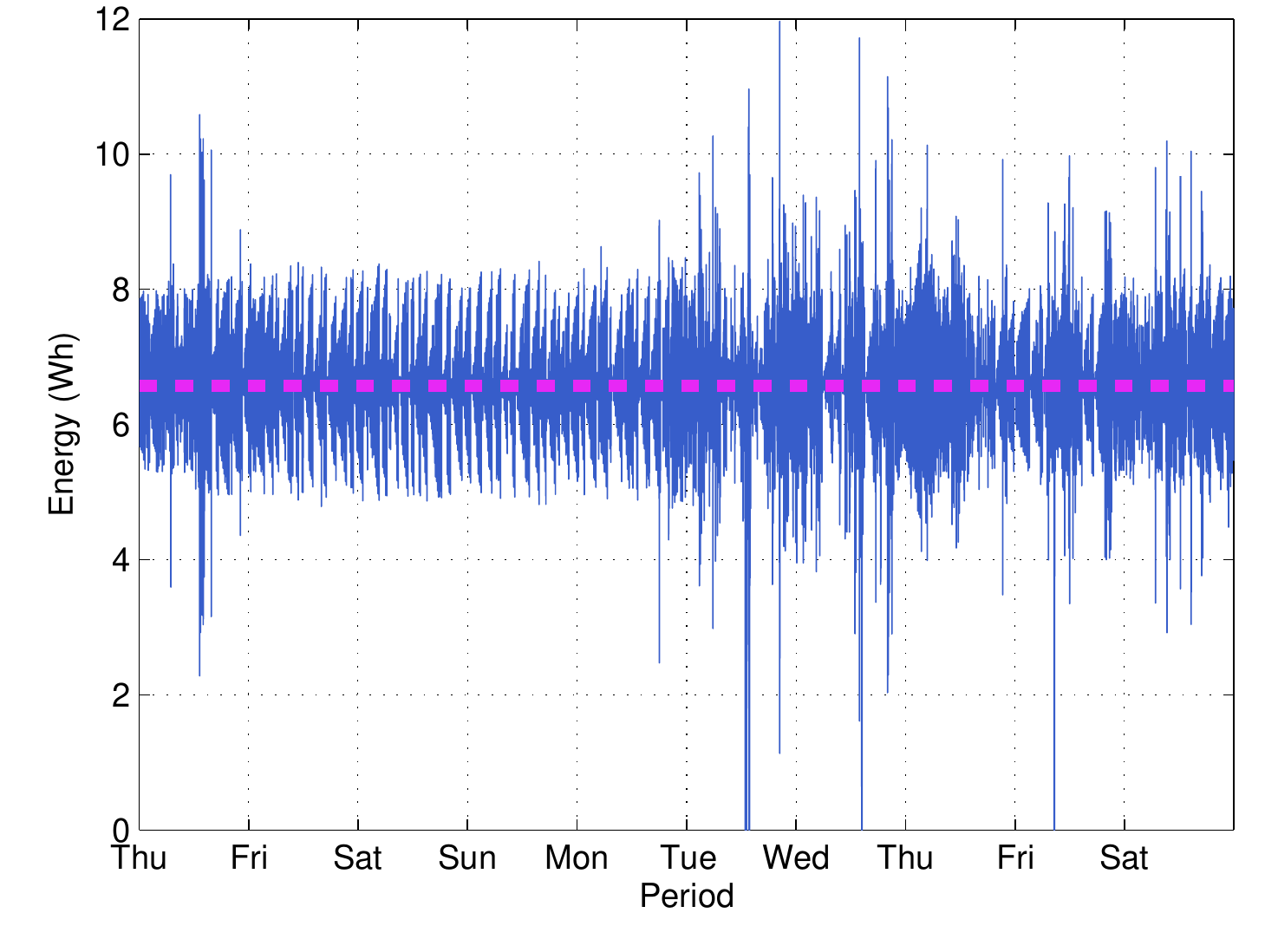}
	\caption{\label{fig:tips:2}Measured energy absorbed by the TV + decoder of site S7. The mean value is also shown.}
\end{figure}

Standby mode is responsible for a relevant energy waste, resulting from an idle status in which the device is neither being operated nor switched off.
The main reason is to ensure prompt reaction upon user request. 
This includes consumer electronics such as DVD players, radio, televisions, as well as air conditioning systems, computers, unplugged phone chargers, etc.
Power consumption due to standby mode may be as low as some mW, but it can exceed tens of W\footnote{Lawrence Berkeley National Laboratory, ``Standby Power Summary Table'', online: http://standby.lbl.gov/summary-table.html}.
The analysis on the GREEND reveals that, in site S7 the television and the decoder are always on in stand-by mode.
Fig. \ref{fig:tips:2} shows the measured energy consumption during subsequent days.
The power consumption is approximately 6.57~W, which yields an annual consumption of 57.57~kWh, equivalently the 1.4\% of the total of site S7 (4099~kWh).
Assuming 10 devices in stand-by mode (e.g., the washing machine, the air conditioning system, the televisions) about 14\% of the total consumption is wasted in standby.

Another device being switched on all the time is the ADSL modem, although users tend to surf the internet only for a few hours a day.
The power consumption of an ADSL modem with WiFi and Ethernet functionalities is about 30~W \footnote{goo.gl/IxWTiO} that means approximately 263~kWh per year.
If we assumed to switch on the modem for about 3 hours per day and during the entire weekend, the total power consumption would be 98~kWh per year, i.e., about 37\% of the energy consumed if the device was always switched on.
Clearly, this applies to all cases in which the connection is not required for other applications/needs, such as VoIP. 
\subsection{Device shifting}\label{sec:shifting}
\begin{figure}[h]
	\centering
	\includegraphics[width= 0.6\columnwidth]{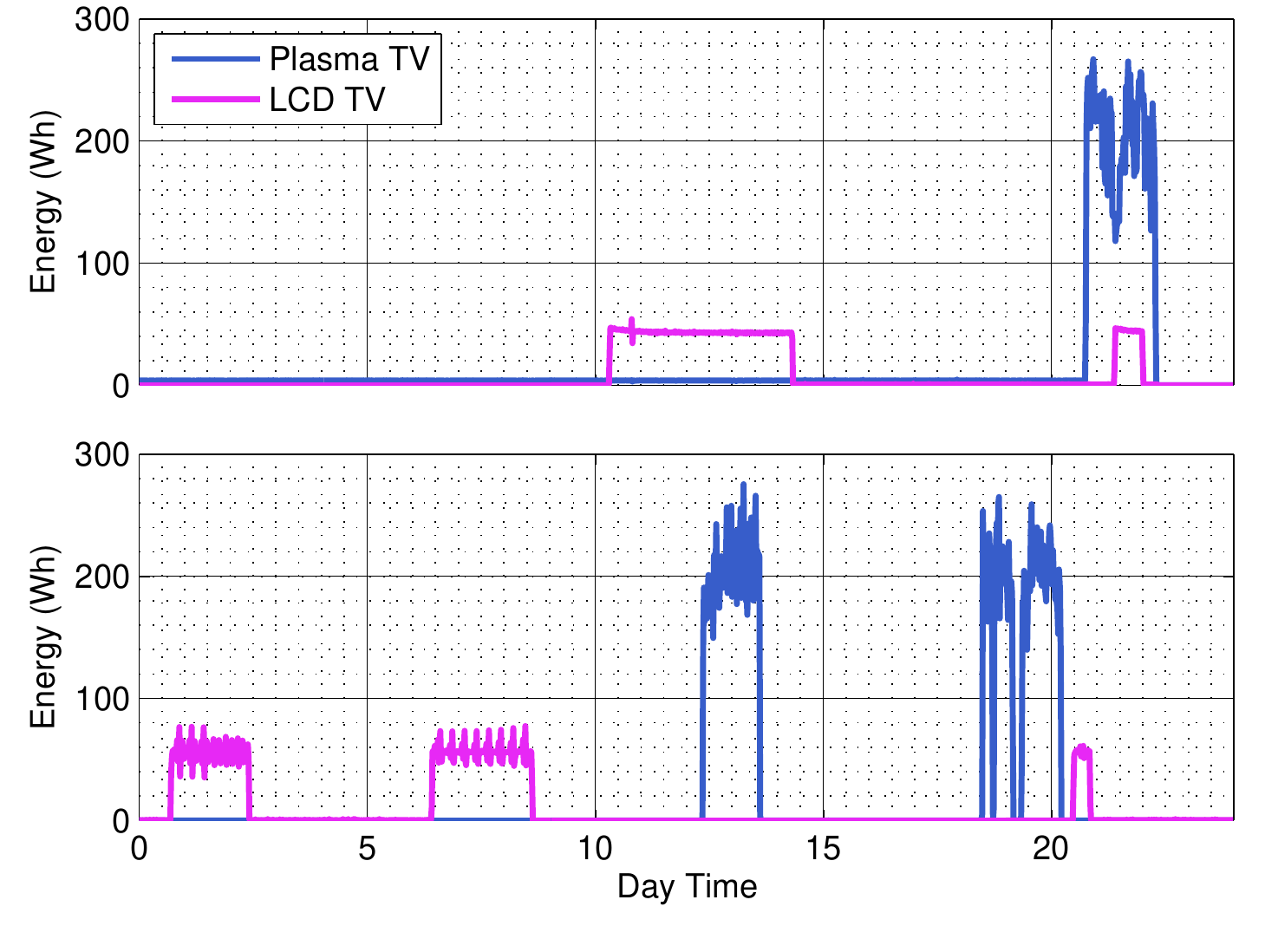}
	\caption{\label{fig:tips:1}Energy consumption comparison of the plasma and the LCD TVs of site S4 (on top) and site S5 (on bottom).}
\end{figure}

Device shifting includes both temporal postponement and improvement of inefficient behavior, namely by reducing the time inefficient devices are used.
To quantify the impact of device shifting we take as example the case of plasma televisions.
Fig. \ref{fig:tips:1} shows the energy consumption profile of the plasma (42'') and the LCD TV (37'') of respectively site S4 and S5 for one day.
As visible, the plasma TV consumes between two and three times the energy of the LCD TV.
Thus a possible way to save energy is to swap the plasma with the LCD television, meaning that the plasma can be used in those rooms whose occupancy is associated to off-peak periods (e.g., bedroom).
Based on Fig.~\ref{fig:tips:1} we can estimate the hourly energy consumption of the plasma and the LCD TV for an hour of activity to be respectively 200~Wh/hour and 80~Wh/hour.
Now the hours of activities for such devices can be obtained as the ratio between the consumed energy and the hourly energy consumption.
For instance, this translates for site S4 as respectively 421 and 148 working hours for the plasma TV and the LCD TV.
In site S5, we estimate 771 and 404 working hours.
Shifting the plasma to off-peak periods can be actuated simply by swapping position of the devices.
This results into 34\% and 23\% lower energy consumption for site S4 and S5, respectively. 
%

\section{Conclusions and future work}\label{sec:conclusions}
In this paper, we investigated the possibility to inform users on energy usage, in order to promote an optimal use of local resources.
In particular, we analysed data collected in several households in Italy and Austria to gain insights into usage behavior.
The main outcome was the formulation of policies to improve energy efficiency in domestic settings and the design of an energy management system, consisting of hardware measurement units and a management software. 
The Mjölnir framework, which we release for open use, provides a platform where various feedback concepts can be implemented and assessed.
This includes widgets displaying disaggregated and aggregated consumption information, as well as daily production and tailored advices.
The formulated policies were implemented as an advisor widget able to autonomously analyze usage and provide tailored energy feedback.
%
While we provided an estimation of potential savings, experiments with end users in real settings will be necessary to ultimately validate the proposed policies.
Further development is also necessary. In particular, we envision the introduction of the following functionalities:
\begin{itemize}
  \item 
  the proposed policies were formulated through the analysis of the GREEND dataset, which while complete for the Austrian and Italian scenarios lacks of generality.
  Thus, the analysis should be extended to include production from renewables and electrical vehicles. New efficiency policies should be introduced accordingly. 
  \item 
  As shown, Mjölnir provides an easy integration between appliance-level metering and external loads disaggregation modules by means of a RESTful interface~\cite{Fielding:2000}.
  We envision the future integration of a load disaggregation unit in the Mjölnir project.
  In particular, we are currently adapting a design based on particle filtering, as in \cite{Egarter2014}.
  \item 
  Knowledge of the likelihood associated to the use of specific devices can improve the effectiveness of advicing for device shifting.
  \item 
  To effectively suggest device shedding, such as for stand-by losses, a model of occupancy should be introduced and automatically learned, as in \cite{Chen2013}.
  \item 
  Beside providing detailed information of energy usage, the benefits of demand response can be exploited only when energy-involving processes can be automated. 
  Therefore solutions are needed to provide a bidirectional communication between controllable loads and the visualization dashboard.
  \item 
  To further develop the set of social features, a comparison of users will be provided.
  In particular, a new widget will be introduced to show the distance from the best and worst performing friend.
\end{itemize}
\section{Acknowledgements}
This work is supported by Lakeside Labs, Klagenfurt, Austria and funded by the European Regional Development Fund (ERDF) and the Carinthian Economic Promotion Fund (KWF) under grant KWF 20214 - 23743 - 35469 and 35470.
The work of F. Versolatto and A. M. Tonello is cofunded by the Interreg IV Italy-Austria ID-6462 CUP B29E12002190007 MONERGY project.

\bibliographystyle{abbrv}
\bibliography{bib.bib}

\end{document}